\documentclass[10pt,fleqn,twocolumn,aps,pre,groupedaddress,a4paper,floatfix]{revtex4}

\usepackage{amsmath}
\usepackage{graphicx,psfrag}
\usepackage{multirow}
\usepackage{graphics}
\usepackage{amsfonts}
\usepackage{amssymb}
\usepackage{subfigure}
\usepackage{natbib}

\begin{document}
 \title{Environment-induced two-mode entanglement in quantum Brownian motion}
 \author{Christian H\"orhammer}
 \email{christian.hoerhammer@uni-bayreuth.de}
 \affiliation{Theoretische Physik I, Universit\"at Bayreuth, D-95440 Bayreuth, Germany}
 \author{Helmut B\"uttner}
 \affiliation{Theoretische Physik I, Universit\"at Bayreuth, D-95440 Bayreuth, Germany}
 \date{\today}
 \begin{abstract}
The time evolution of quantum correlations of entangled two-mode continuous variable states is examined in single-reservoir as well as two-reservoir models, representing noisy correlated or uncorrelated non-Markovian quantum channels. For this purpose the model of quantum Brownian motion is extended. Various separability criteria for Gaussian continuous variable systems are applied. In both types of reservoir models moderate non-Markovian effects prolong the separability time scales. However, in these models the properties of the stationary state may differ. In the two-reservoir model the initial entanglement is completely lost and both modes are finally uncorrelated. In a common reservoir both modes interact indirectly via the coupling to the same bath variables. Below a critical bath temperature entanglement between the two modes is preserved even in the steady state. A separability criterion is derived, which depends on the bath temperature and the response function of the open quantum system. Thus, the extended quantum Brownian motion model of a two-mode continuous variable system in a common reservoir provides an example of environment-induced entanglement.
\end{abstract}
\maketitle
\section{Introduction}
During the last decade quantum information and computation has been extended from discrete systems to quantum systems with continuous variables such as position and momentum or the amplitudes of electromagnetic field modes. This quantum information theory of continuous variable systems has received much attention in the past few years \cite{braunstein2, braunstein, cerf} and has found various applications in quantum cryptography and quantum teleportation \cite{furusawa, kimble}. In this context, Gaussian states play a prominent role since they can be easily created and controlled experimentally and are less affected by decoherence. Great advances have been made in characterizing the entanglement properties of two-mode Gaussian states by determining the necessary and sufficient criteria for their separability \cite{simon, duan} and by developing quantitative entanglement measures \cite{vidal, wolf}. The prototype of these states are the two-mode squeezed states, which have been successfully produced via nonlinear parametric down conversation and applied to quantum teleportation \cite{ou, zhangtc}.\\

Due to the unavoidable interaction with the environment, any pure quantum state used in some quantum information process evolves into a mixed state. Thus, a realistic analysis of continuous variable quantum channels must take decoherence and dissipation into account. Within the theory of open quantum systems \cite{breuer, dittrich} the dissipative dynamics are mainly described by master equations of the reduced density matrix. Initial quantum superpositions are destroyed and quantum correlations are lost during characteristic decoherence and separability time scales. The Markovian time evolution of quantum correlations of entangled two-mode continuous variable states has been examined in single-reservoir \cite{prauzner, braun} and two-reservoir models \cite{duan, halliwell, olivares2, serafini2}, representing noisy correlated or uncorrelated Markovian quantum channels. Quantum correlations are found to be better preserved in a common reservoir. Additionally the coupling to the same bath variables might generate new quantum correlations between the parts of the subsystem. This effect of environment-induced entanglement has already been studied for discrete systems \cite{braun, kim, yiyu} and can lead to asymptotically entangled states \cite{plenhuelgo, benatti1, benatti2, rajagopal2}. The underlying Born-Markov Approximation assumes weak coupling between the system and the environment to justify a perturbative treatment and neglects short-time correlations between the system and the reservoir. This approach has been widely and successfully employed in the field of quantum optics \cite{walls} where the characteristic time scales of the environmental correlations is much shorter compared to the internal system dynamics. Challenged by new experimental evidence a growing interest in non-Markovian descriptions can be observed. Very recently some phenomenological \cite{ban2, mcaneney} and microscopic models \cite{manisc6, an, liu, an2} of non-Markovian quantum channels have been proposed. Using the analogy between the Hilbert space of quantized electromagnetic fields and the Hilbert space of quantum harmonic oscillators, the Caldeira Leggett model of quantum Brownian motion \cite{caldeira3, ford3, paz} can be extended to describe the entanglement dynamics of two-mode squeezed states.\\

In this paper, the time evolution of quantum correlations of initially entangled two-mode continuous variable states is numerically examined in a common reservoir model. The focus is on non-Markovian influences and strong coupling effects. The non-Markovian dynamics are described by an extended two-mode version of Hu-Paz-Zhang master equation of quantum Brownian motion. In the single or common reservoir model both oscillators (modes) are coupled to the same reservoir variables, whereas in the two-reservoir model each oscillator (mode) is interacting with its own independent reservoir. In both types of reservoir models moderate non-Markovian effects prolong the separability time scales which also depend on the interaction strength between the system and the environment. However, in these models the properties of the stationary state may differ. In the two-reservoir model the initial entanglement is completely lost and both modes are finally uncorrelated. In a common reservoir both modes interact indirectly via the coupling to the same bath variables. Therefore, new quantum correlations may emerge between the two modes. Below a critical bath temperature entanglement is preserved even in the steady state. A separability criterion is derived, which depends on the bath temperature and the response function of the open quantum system. Thus, the extended quantum Brownian motion model of a two-mode continuous variable system in a common reservoir provides an example of environment-induced quantum two-mode entanglement.\\

The paper is organized as follows. In section II, we briefly review various separability criteria for two-mode Gaussian states and the Markovian separability times for two-mode squeezed states in single- and two-reservoir models. In section III we shortly describe the Hu-Paz-Zhang master equation of quantum Brownian motion which is the basis for studying non-Markovian effects. We resume the extended, two-mode-version of the Caldeira Leggett model for single and two-reservoir models and introduce a modified common-reservoir model.  In section IV we present and discuss the numerical results of the entanglement dynamics of two mode squeezed states in the modified common-reservoir model. Different scenarios are analyzed, including the case of noise-induced steady state entanglement. A simple separability criteria for the stationary two-mode state is derived. Finally, a brief summary is given in section V.
\section{Continuous Variable Systems}

\subsection{Separability criteria for two mode Gaussian states}
In the following we review the separability criteria for a special class of continuous variable systems -- the two mode Gaussian states. A Gaussian two mode state with coordinates  $q_1$, $q_2$ und momenta $p_1$, $p_2$ has a Gaussian Wigner function in semi-classical phase space
\begin{equation}
W({\mathbf X})=\frac{1}{4\pi^2\sqrt{\det {\mathbf
V}}}\exp\left[-\frac{1}{2}\mathbf {X V^{-1} X^T}\right]
\end{equation}
with $\mathbf{X}=(q_1,p_1,q_2,p_2)$ and commutation relation $[q_i,p_j]=i\hbar\delta_{ij}$ for $i,j=1,2$. It is completely characterized by its first and second moments. \\The $4\times4$ - covariance matrix
\begin{equation}\label{KovarianzV}
\mathbf{V}=\left(\begin{array}{cc}A&C\\C^T&B\end{array}\right),\quad V_{ij}=\frac{1}{2}\langle X_i X_j+X_jX_i\rangle
\end{equation}
(where all first moments $\langle X_j\rangle$ have been set to zero) contains four local symplectic invariants in form of the determinants of the block-matrices $\mathbf A$, $\mathbf B$, $\mathbf{C}$ and $\mathbf V$. A Gaussian continuous variable state is separable if and only if the partial transpose $\tilde\rho$ of its density matrix $\rho$ is non-negative (PPT-criterion). Based on the above invariants Simon \cite{simon} has derived a PPT-criterion for bipartite Gaussian continuous variable states,
\begin{equation}\label{simkrit}
S(t)=\det\mathbf{V}-\frac{1}{4}\left(\det\mathbf{A}+\det\mathbf{B}+2|\det\mathbf{C}|\right)+\frac{1}{16}\geq 0
\end{equation}
which is also a necessary separability criterion for non-Gaussian states. The PPT-criterion has a geometrical interpretation as mirror reflection of the Wigner function in phase space. In case of a Gaussian two mode state the partial transpose coincides with a change of the signs in those elements of the covariance matrix, which connect the momentum of the first mode to the coordinate of the second mode. Equivalent to that is the criterion  $\tilde \nu_-\geq\frac{1}{2}$ for separability where $\tilde \nu_{\pm}$ are the symplectic eigenvalues of the partial transposed density matrix $\tilde\rho$:
\begin{eqnarray}\label{nupm}
\tilde \nu_{\pm}=\frac{1}{\sqrt{2}}\left[\tilde\Delta_V\pm\sqrt{\tilde\Delta_V^2-4\det
V}\right]^{\frac{1}{2}}
\end{eqnarray}
with $\tilde\Delta_V=\det A+\det B-2\det C$. With the symplectic eigenvalue $\tilde \nu_{-}$ the logarithmic negativity as a quantitative measure of entanglement can be defined by \cite{vidal, adesso}:
\begin{eqnarray}
\label{logNeg} E_{\mathcal
N}(\rho)&=&\mbox{max}\left\{0,-\ln2\tilde\nu_-\right\}.
\end{eqnarray}
In order to apply the PPT-criterion the complete knowledge of all second moments is required. For practical purposes there are also weaker separability criteria based on linear or quadratic combinations of only a few elements of the covariance matrix. Duan et. al. \cite{duan} derived such a criterion.  Starting point is the definition of a pair of EPR-like operators $\hat u=|a|\hat q_1+\frac{1}{a}\hat q_2$ and $\hat v=|a|\hat p_1-\frac{1}{a}\hat p_2$ with $a\in\mathbb{R}\setminus\{0\}$
fulfilling the commutation relations $[\hat q_k,\hat
p_j]=i\delta_{kj}$, $j,k=1,2$. Then, for every separable bipartite quantum state $\rho$, the sum of the variances fulfills the relation
\begin{equation}\label{eprsep}
\langle (\Delta \hat u)^2\rangle_{\rho}+\langle (\Delta \hat v)^2\rangle_{\rho}
\geq a^2+\frac{1}{a^2}
\end{equation}
For Gaussian states this set of inequalities (for all possible values of $a$) completely characterizes the set of separable states and is equivalent to eq. \eqref{simkrit}. The product version of this inequality for $a=1$ is given by \cite{reid2, tan, giovannetti}
\begin{equation}\label{eprtan}
\langle (\Delta \hat u)^2\rangle_{\rho}\cdot\langle (\Delta \hat
v)^2\rangle_{\rho} \geq 1,
\end{equation}
which is a special case of the generalized curved quadratic entanglement witnesses \cite{eisert4}. In general, the product witnesses are stronger tests than the respective linear tests. Thus, the set of entangled covariance matrices detected by this quadratic test is strictly larger than that detected by linear combinations of second moments \cite{eisert4}. \\
We also will apply a classification scheme of quantum states based on marginal and global purities of a bipartite Gaussian quantum system whereas $\mu=\mbox{Tr}[\rho^2]$ is the purity of the total system and $\mu_1=\mbox{Tr}[\rho_1^2]$, $\mu_2=\mbox{Tr}[\rho_2^2]$ are the purities of the reduced density matrices $\rho_1=\mbox{Tr}_2[\rho]$, $\rho_2=\mbox{Tr}_1[\rho]$ of the two-mode system. According to ref. \cite{adesso3, adesso2}, a two-mode Gaussian mixed state is separable if
\begin{equation}\label{kritpurities}
\mu_1\mu_2\leq\mu\leq\frac{\mu_1\mu_2}{\mu_1+\mu_2-\mu_1\mu_2},
\end{equation}
and it is entangled for the case that
\begin{equation}\label{kritpur2}
\frac{\mu_1\mu_2}{\sqrt{\mu_1^2+\mu_2^2-\mu_1^2\mu_2^2}}<\mu\leq\frac{\mu_1\mu_2}{\mu_1\mu_2-|\mu_1-\mu_2|}.
\end{equation}
In between there is a coexistence region
\begin{equation}
\frac{\mu_1\mu_2}{\mu_1+\mu_2-\mu_1\mu_2}<\mu\leq\frac{\mu_1\mu_2}{\sqrt{\mu_1^2+\mu_2^2-\mu_1^2\mu_2^2}},
\end{equation}
where both separable and entangled states can be found. The calculation of the global and marginal purities thus provides analytical bounds on the entanglement of Gaussian states.

\subsection{Markovian separability times for two-mode squeezed states}
The separability criterion \eqref{simkrit} is simplified if the covariance matrix $V$ is transformed to standardform by symplectic transformations:
\begin{equation}\label{Vtwomode}
V_{\rm st}=\left(\begin{array}{cccc}a&0&c_+&0\\ 0&a&0&c_-\\
c_+&0&b&0\\ 0&c_-&0&b\end{array}\right),
\end{equation}
where the elements $a$, $b$, $c_+$ and $c_-$ of
$V_{\rm st}$ are determined by the four symplectic invariants
$\det \mathbf{A}=a$ $\det \mathbf{B}=b$,  $\det\mathbf{V}=(ab-c_+^2)(ab-c_-^2)$ and $\det \mathbf{C}=c_+c_-$. In a Markovian two-reservoir model the dynamics of the two-mode state are described by the quantum optical master equation of the damped quantum oscillator (interaction picture)
\begin{equation}
\dot \rho=\sum\limits_{j=1}^2\frac{\gamma}{2}\left(\bar{n}L[a_j^{\dagger}]\rho+(\bar{n}+1)L[a_j]\rho\right),
\end{equation}
with mean bosonic occupation number $\bar{n}=(e^{\beta\hbar\omega_0}-1)^{-1}$, damping constant $\gamma$ and $L[o]\rho=2o\rho o^{\dagger}-o^{\dagger}o\rho-\rho
o^{\dagger}o$. The time evolution of the matrix elements is then given by $a(t)=ae^{-\gamma t}+\Delta(t)$,  $b(t)=be^{-\gamma t}+\Delta(t)$ and $c_{\pm}(t)=\pm c_{\pm}e^{-\gamma t}$ with $\Delta(t)=(2\bar{n}+1)(1-e^{-\gamma t})$ \cite{olivares2}.
In the case of a two mode squeezed vacuum state
\begin{equation}
|\psi_{\xi}\rangle=e^{\xi^* \hat
a_1\hat a_2-\xi \hat a_1^{\dagger}\hat a_2^{\dagger}}|0\rangle_1|0\rangle_2
\end{equation}
with squeezing parameter $\xi$, the initial covariance matrix is already given in standard form with elements
$a=b=\frac{1}{2}\cosh(2|\xi|)$ and $c_{\pm}=\pm\frac{1}{2}\sinh(2|\xi|)$. Applying the above separability criterion, a Markovian separability time can be derived from the separability function \eqref{simkrit} which is reduced to $S(t)=e^{-2|\xi|}e^{-\gamma t}+\Delta(t)-1$. The time $\tau_1$ after which the initial entanglement is lost is given by the condition $S(\tau_1)=0$ and reads
\begin{equation}\label{tsep}
\tau_{1}=\frac{1}{\gamma}\ln\left(1+\frac{1-e^{-2|\xi|}}{2\bar{n}}\right)
\end{equation}
\cite{duan, olivares2, serafini2} for the two-reservoir model yielding $\tau_{1}\to\infty$ for $\bar{n}\to0$.
In the case of a single reservoir model where the corresponding master equation contains additional terms, the time evolution of the covariance matrix elements is slightly different and results in the separability time \cite{prauzner}
\begin{equation}
\tau_{2}=\frac{1}{2\gamma}\ln\left(\frac{2\bar{n}+1-e^{-2|\xi|}}{2\bar{n}+1-e^{2|\xi|}}\right).
\end{equation}
Thus a common reservoir extends the Markovian separability time. Furthermore the initial entanglement is partially preserved even for $\bar{n}>0$ if the initial squeezing exceeds a critical value $ \xi_c=\frac{1}{2}\ln\left(2\bar{n}+1\right)$. These results change if the dynamics are described without the Born-Markov assumption.
\section{Non-Markovian Entanglement Dynamics}
\subsection{HPZ master equation of quantum Brownian motion}
Non-Markovian effects are discussed here on the basis of the Caldeira-Leggett model of quantum Brownian motion \cite{caldeira1,
caldeira2, caldeira3} often referred to as independent-oscillator-model \cite{ford4, ford3}. It is a system plus reservoir model where the total Hamiltonian consists of three parts
\begin{equation}
H=H_s+H_b+H_{\rm int},
\end{equation}
with $H_s$ as Hamiltonian of the subsystem which interacts via the Hamiltonian $H_{\rm int}$
with a bath that is described by a collection of a large number of harmonic oscillators $H_b=\sum_i\hbar\omega_i(b^{\dagger}b+1)$.
In detail the Hamiltonian of the Caldeira Leggett model is given by
{\mathindent0.5cm \begin{equation}
\label{Hcl}H=\frac{p^2}{2m}+V(q)+\sum_{i=1}^N\left[\frac{p_i^2}{2m_i}+\frac{m_i\omega_i^2}{2}
\left(x_i-\frac{c_iq}{m_i\omega_i^2}\right)^2\right],
\end{equation}}
where $q$ and $p$ are the Heisenberg-operators of coordinate and momenta of the Brownian oscillator moving in an harmonic potential $V(q)=\frac{1}{2}m\omega_0^2q^2$ and coupled to a bath of $N$ independent harmonic oscillators with variables $x_i$, $p_i$ and frequencies $\omega_i$. The bath is characterized by its spectral density
\begin{equation}\label{Jdrude}
J(\omega)=\pi\sum_{i=1}^N\frac{c_i^2}{2m\omega_i}\delta(\omega-\omega_i)=\frac{\gamma\omega\Gamma^2}{\omega^2+\Gamma^2}.\end{equation}
The interaction is bilinear in the coordinates $q$ and
$x_i$ of the subsystem and the bath respectively with coupling parameters $c_i$. The self-interaction term
(proportional to $q^2$) in the Hamiltonian
\begin{equation}\label{Hint}
H_{\rm
int}=\sum_i\left[-c_ix_iq+{c_i^2\over2m_i\omega_i^2}q^2\right]
\end{equation}
renormalizes the oscillator potential to ensure that the observable frequency is close to bare oscillator frequency $\omega_0$. From influence functional path integral techniques Hu, Paz, Zhang have derived the master equation
{\mathindent0.5cm\begin{eqnarray}\label{hpzeq}
\dot\rho&=&\frac{1}{ i\hbar
}\left[H_s,\rho\right]+\frac{m\delta\Omega^2(t)}{2i\hbar}[q^2,\rho]+\frac{\gamma_p(t)}{2i\hbar}[q,\{p,\rho\}]\nonumber\\
& &+\frac{D_{qp}(t)}{\hbar^2}[q,[p,\rho]]-\frac{D_p(t)}{\hbar^2}[q,[q,\rho]],
\end{eqnarray}}
with $[\,.\,]$ and $\{\,.\,\}$ denoting commutator and anti-commutator respectively. This master equation is valid for arbitrary coupling and temperature. The non-Markovian character is contained in the time-dependent coefficients which read in expansion up to the second order in the system-bath coupling constant \cite{paz}:
{\mathindent0.5cm
\begin{eqnarray}\label{hpzkoeff1}
\gamma_p(t)&=&\frac{2}{\hbar
m\omega_0}\int_0^tdt'L(t')\sin\omega_0t',\\
\label{hpzkoeff2}
\delta\Omega^2(t)&=&\frac{\gamma\Gamma}{m}-\frac{2}{\hbar
m}\int_0^tdt'L(t')\cos\omega_0
t',\\
\label{hpzkoeff3}
D_{qp}(t)&=&\frac{1}{m\omega_0}\int_0^tdt'K(t')\sin\omega_0t',\\
\label{hpzkoeff4}
D_p(t)&=&\int_0^tdt'K(t')\cos\omega_0t',
\end{eqnarray}}
where $L(t)=i\langle [\eta(t),\eta(0)]\rangle$ and $K(t)=\frac{1}{2}\langle\{\eta(t),\eta(0)\}\rangle$ are connected to the spectral density \eqref{Jdrude} by
\begin{eqnarray}
L(t)&=&\frac{\hbar}{\pi}\int_0^{\infty}d\omega\,
J(\omega)\sin\omega t,\\
K(t)&=&\frac{\hbar}{\pi}\int_0^\infty d\omega\,
J(\omega) \coth(\frac{1}{2}\beta\hbar\omega)\cos\omega t.
\end{eqnarray}
$K(t)$ is the correlation function of the quantum noise term $\eta$ resulting from averaging over the initial thermal bath distribution. The exact expressions of the HPZ-coefficients are related to the Green's functions of the corresponding quantum Langevin equations \cite{karrlein, reibold}. The entanglement properties of the joint state of the oscillator and its environment have been studied in ref. \cite{eisert}.
\subsection{Two-reservoir model}
The dynamics of two identical, not directly interacting modes (with coordinates and momenta $q_j,p_j$, $j=1,2$) in two uncorrelated reservoirs is modeled by the interaction Hamiltonian
\begin{equation}
H_{\rm int}=-q_1\sum\limits_{i=1}^{\infty}c_ix_i^b-q_2\sum\limits_{i=1}^{\infty}c_ix_i^c
\end{equation}
with $\langle x_i^bx_j^c+x_i^cx_j^b\rangle=0\quad\forall\, i,j$. The master equation of the reduced density matrix is then given by the sum of the master equations of two single modes \cite{manisc6}:
\begin{eqnarray}\label{hpzeq2bath}
\dot\rho&=&\sum_{j=1}^2\left\{\left[\frac{p_j^2}{2i\hbar
m}+\frac{m\gamma_q(t)q_j^2}{2i\hbar
},\rho\right]+\frac{\gamma_p(t)}{2i\hbar}[q_j,\{p_j,\rho\}]\right.\nonumber\\
&&+\left.\frac{D_{qp}(t)}{\hbar^2}[q_j,[p_j,\rho]]-\frac{D_p(t)}{\hbar^2}[q_j,[q_j,\rho]]\right\}.
\end{eqnarray}
The time dependent coefficients are given by $\gamma_q(t)=\omega_0^2+\delta\Omega^2(t)-\gamma\Gamma/m$
and eq. \eqref{hpzkoeff1} to \eqref{hpzkoeff4}. The time evolution of a two-mode squeezed state in two uncorrelated non-Markovian channels has been studied recently in ref. \cite{manisc6}. The authors derived a non-Markovian separability function which shows oscillations in case of an artificial {\sl out of resonance} bath with $\Gamma\ll\omega_0$. In this two-reservoir model the initial entanglement is always completely lost and both modes are finally uncorrelated (even at zero temperature while $\tau_1\to\infty$). This may not be the case in a common reservoir model as will be shown in the next sections.

\subsection{Common-reservoir model}
The dynamics of two identical, not directly interacting modes in a common reservoir is modeled by the interaction Hamiltonian
\begin{equation}\label{HWW1b}
H_{\rm int}=-q_1\sum\limits_{i=1}^{\infty}c_iq_i-q_2\sum\limits_{i=1}^{\infty}c_iq_i.
\end{equation}
The corresponding master equation for a two-mode system with $H_s=H_{s1}+H_{s2}$ is then given by \cite{hu2mode}
\begin{eqnarray}\label{hpzeq1bath}
\dot\rho&=&\frac{1}{ i\hbar
}\left[H_s,\rho\right]+\frac{M\delta\tilde\Omega^2(t)}{i\hbar}[R^2,\rho]+\frac{\gamma_p(t)}{i\hbar}[R,\{P_R,\rho\}]\nonumber\\
&&+\frac{2D_{qp}(t)}{\hbar^2}[R,[P_R,\rho]]-\frac{2D_p(t)}{\hbar^2}[R,[R,\rho]]
\end{eqnarray}
with $\delta\tilde\Omega^2(t)=\delta \Omega^2(t)-\gamma\Gamma/m$. As one can see immediately, the master equation  \eqref{hpzeq1bath} contains only the normal coordinates $R=(q_1+q_2)/2$ and $P_R=p_1+p_2$ but not the relativ coordinate $x=q_1-q_2$. This results from the specific coupling $\sim x_i(q_1+q_2)$ and from the choice of two identical modes with $\omega_1=\omega_2$. In this case, the motion of the relativ coordinate is not accompanied by dissipation \cite{hoebu, caldeira2oszi}. Additional assumptions are therefore necessary to describe the relaxation dynamics completely.

\subsection{Modified common-reservoir model}
The Hamiltonian \eqref{HWW1b} describes a bilinear coupling in the coordinates of the two modes and the coordinates of the bath variables. The subsystem is thus effectively coupled by the center-of-mass $R=\frac{1}{2}(q_1+q_2)$. The motion of the relative coordinate however is unitary. Therefore, in the case of an initial two mode squeezed state only half of the entanglement is lost in the stationary state regardless of the coupling strength \cite{an2}. In order to be able to model a complete loss of the initial entanglement, an additional assumption about the dynamics of the relative coordinate is necessary if the two modes are identical. To be able to model the dissipative dynamics for arbitrary coupling we choose an interaction Hamiltonian similar to the original Caldeira-Leggett model substituting the variable $q$ by $R$. A renormalization term proportional to $R^2$ ensures that the frequency shift is compensated in the stationary state. The modified interaction Hamiltonian thus reads
\begin{equation}\label{HintR}
H_{\rm
int}=-\sum\limits_{i=1}^{\infty}c_ix_iR+
\sum\limits_{i=1}^{\infty}\frac{c_i^2}{2m_i\omega_i^2}R^2
\end{equation}
to guarantee that the eigenvalue equation for the eigenfrequencies $\nu$ of the total system
\begin{equation}
\left[\nu^2-\omega_1^2+h(\nu)\right]\left[\nu^2-\omega_2^2+h(\nu)\right]-h^2(\nu)=0
\end{equation}
with $h(\nu)=\sum_i\frac{c_i^2\nu^2}{4\omega_i^2(\omega_i^2-\nu^2)}$, has just eigenvalues $\nu\geq0$ (all masses $m_i$ have been set to one). The frequency shift in the stationary state is compensated and the observable and the bare oscillator frequency coincide. The Hamiltonian of the total system consists of the Hamiltonian of the two modes $H_s=H_{s1}+H_{s2}$, the Hamiltonian  $H_b=\sum_i\hbar\omega_i(b^{\dagger}b+1)$ of the bath modes as well as the interaction Hamiltonian \eqref{HintR}. The system of coupled Heisenberg equations of motion can be solved equivalently to the case of a single Brownian oscillator and leads to two coupled quantum Langevin equations for the two modes $j=1,2$:
\begin{equation}\label{Glq1q2}
\ddot
q_j=-\omega_j^2q_j-\frac{\gamma(t) R(0)}{M}-\int_0^t dt'\frac{\gamma(t-t')\dot
R(t')}{M}+\frac{\eta(t)}{M},
\end{equation}
with $\gamma(t)=\gamma\Gamma e^{-\Gamma t}$. In the case of two identical modes with $\omega_j=\omega_0$ for $j=1,2$ and total mass $M=2m$ the system of coupled Langevin equations \eqref{Glq1q2} reduces to a quantum Langevin equation of the center-of-mass motion
\begin{equation}\label{langevinR} \ddot R= -\omega_0^2R- \frac{\gamma(t)R(0)}{M}-
\int_0^t{dt'\frac{\gamma(t-t') \dot R(t')}{M}} +\frac{\eta(t)}{M},\end{equation}
with stationary correlations
{\mathindent0.5cm\begin{eqnarray}
\label{xxchi}
\langle R^2\rangle&=&\frac{\hbar}{\pi}\int_{0}^{\infty}d\omega\, \coth(\frac{1}{2} \beta \hbar \omega)\,
\mbox{Im}\{\tilde\chi(\omega)\},\\
\label{ppchi}
\langle P_R^2\rangle&=&\frac{\hbar}{\pi}\int_{0}^{\infty}d\omega\, M^2\omega^2 \coth(\frac{1}{2} \beta \hbar \omega)\,\mbox{Im}\{
\tilde\chi(\omega)\},\end{eqnarray}}
where $\tilde\chi(\omega)=\left[M\omega_0^2-M\omega-i\omega\tilde\gamma(\omega)\right]^{-1}$ with $\tilde \gamma (\omega)=\int_0^{\infty}dt\, \gamma(t)e^{i\omega t}$ is the susceptibility of the non-Markovian damped harmonic oscillator.
The motion of the relative coordinate $x=(q_1-q_2)$ is decoupled from the motion of $R$. In order to study the case of two identical modes with dissipation in the relative motion additional assumptions make sense. It seems plausible to assume a weakly dissipative dynamics of the relative coordinate given by the solution of the Born-Markovian master equation of the damped harmonic oscillator in the form
\begin{equation}\label{xrelax}
\langle x^2(t)\rangle=\langle x_0^2\rangle e^{-\gamma_p t}+2(1-e^{-\gamma_p t})(2\bar{n}+1),
\end{equation}
where $\gamma_p=\lim_{t\to \infty}\gamma_p(t)$ is the stationary (Markovian) value of the coefficient \eqref{hpzkoeff1}. The initial value $\langle x_0^2\rangle$ in case of a squeezed two mode state is given by  $\langle x_0^2\rangle=e^{-2\xi}$. The dynamics of the normal coordinates is uncorrelated with $\langle \{x,R\}(t)\rangle =0$. Of course, correlations in the original coordinates are still present.
\section{Numerical Analysis of the Modified Common-Reservoir Model}
In this section the non-Markovian entanglement dynamics of a squeezed two mode state in the modified common reservoir model is numerically analyzed.  The initial covariance matrix is given by \eqref{Vtwomode} with elements
$
\langle q_j^2\rangle=\langle p_j^2\rangle=\frac{1}{2}\cosh(2\xi)$ for $j=1,2$ and
$\langle \{q_1,q_2\}\rangle=-\langle \{p_1,p_2\}\rangle=\sinh(2\xi)
$ (with $m=1$, $\omega_0=1$ and $\hbar=1$). The corresponding initial values of the variances of the normal coordinates therefore read $\langle R^2\rangle = \frac{1}{4}e^{2\xi}$, $\langle P_R^2\rangle = e^{-2\xi}$, $\langle x^2\rangle = e^{-2\xi}$, and $\langle p_x^2\rangle = \frac{1}{4}e^{2\xi}$
with minimal uncertainty $\langle R^2\rangle\langle P_R^2\rangle=\frac{1}{4}$ and $\langle x^2\rangle\langle p_x^2\rangle=\frac{1}{4}$.\\

\begin{figure}[t]
\psfrag{y}[c]{\small $E_{\mathcal N}(t)$, $S(t)$, $\mu(t)$} \psfrag{x}[c]{$\omega_0 t$} \psfrag{b}[c]{$\tau_s$}\psfrag{a}[c]{$E_{\mathcal N}$}\psfrag{c}[c]{$S(t)$}\psfrag{d}[c]{$\mu(t)$}
\begin{center}
 \includegraphics[width=7.5cm,clip]{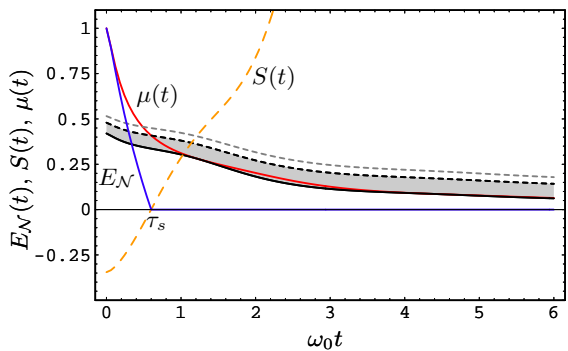}
 \caption[]{\label{fig1}\sl\small Time evolution of the logarithmic negativity $E_{\mathcal N}(t)$ \eqref{logNeg}(blue line), separability function $S(t)$ \eqref{simkrit} (yellow line) and purity $\mu(t)$ (red line). The initial entanglement is lost when $E_{\mathcal N}=0$ and $\mu(t)$ is in the entanglement region \eqref{kritpur2} (shaded grey) respectively. $E_{\mathcal N}=0$ for $t>\tau_s$ where $\tau_s<\tau_1,\tau_2$ (here $\tau_{1}\approx \omega_0^{-1}$ and $\tau_{2}\approx 2.2\omega_0^{-1}$) due to the effect of non-weak interaction strength $\gamma=0.2\omega_0$. Parameters are $|\xi|=1$ and $\Gamma=10\omega_0$, $T=3.5\hbar\omega_0/k$. }
 \end{center}
\end{figure}

\begin{figure}[t]
 \psfrag{y}[c]{\small $E_{\mathcal N}(t)$, $S(t)$, $\mu(t)$} \psfrag{x}[c]{$\omega_0 t$}\psfrag{b}[l]{$\tau_s$}
\begin{center}
 \includegraphics[width=7.5cm,clip]{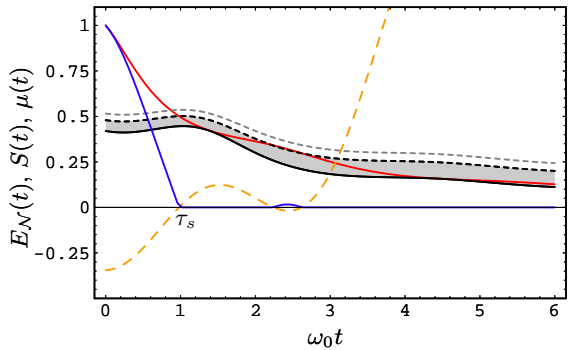}
 \caption[]{\label{fig2}\sl\small Non-Markovian effects (reducing cut-off frequency to $\Gamma\gtrsim\omega_0$) prolong the separability time (compared to $\Gamma\gg\omega_0$) and may lead to a Gaussian decay of the logarithmic negativity. Partial revivals of the logarithmic negativity can occur.  Here $\Gamma=\omega_0$. Parameters $\xi$, $\gamma$, $T$ and colors of the functions $E_{\mathcal N}(t)$, $S(t)$, $\mu(t)$ are chosen as in fig. 1.}
 \end{center}
\end{figure}

\begin{figure}[t]
 \psfrag{y}[c]{\small $E_{\mathcal N}(t)$, $S(t)$, $\mu(t)$} \psfrag{x}[c]{$\omega_0 t$}\psfrag{a}[l]{$\tau_e$}\psfrag{b}[l]{$\tau_s$}\psfrag{c}[l]{$E_{\mathcal{N}}(\infty)$}
\begin{center}
 \includegraphics[width=7.5cm,clip]{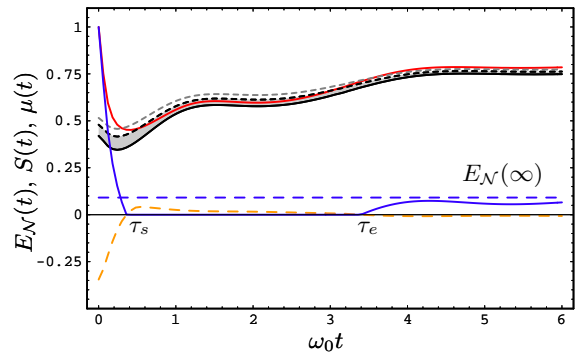}
 \caption[]{\label{fig3}\sl\small At low temperature and strong coupling, the two-mode state can get re-entangled after a certain time $\tau_e$ with asymptotically constant quantum correlations. Parameters here are $\gamma=1.5\omega_0$, $\Gamma=10\omega_0$ and $T=10^{-3}\hbar\omega_0/k$ (with Markovian separability times $\tau_1,\tau_2\to\infty$). Symbols and colors are chosen as in fig. 1.}
 \end{center}
\end{figure}

\begin{figure}[t]
 \psfrag{y}[c]{\small $E_{\mathcal N}(t)$, $S(t)$, $\mu(t)$} \psfrag{x}[c]{$\omega_0 t$}\psfrag{a}[l]{$\tau_e$}\psfrag{b}[l]{$\tau_s$}\psfrag{c}[l]{$E_{\mathcal{N}}(\infty)$}
\begin{center}
 \includegraphics[width=7.5cm,clip]{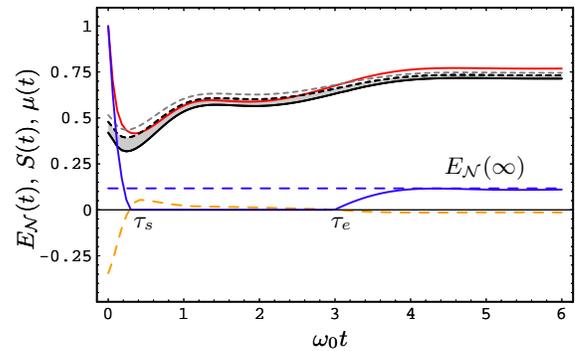}
 \caption[]{\label{fig4}\sl\small Increasing the coupling strength $\gamma$ reduces the separability time $\tau_s$ as well as the re-entangling time $\tau_e$ (compared to fig. 3). Initial entanglement is destroyed faster while asymptotic quantum correlations are generated sooner. Parameters here $\gamma=2\omega_0$, $\Gamma=10\omega_0$ and $T=10^{-3}\hbar\omega_0/k$. Symbols and colors are chosen as in fig. 1. }
 \end{center}
\end{figure}

Figures \ref{fig1}-\ref{fig4} give examples of the time evolution of a two mode squeezed states in a common reservoir, in form of the logarithmic negativity $E_{\mathcal N}(t)$ \eqref{logNeg}(blue line), separability function $S(t)$ \eqref{simkrit} (yellow line) and purity $\mu(t)$ (red line). While for small coupling parameter $\gamma$ and large cut-off frequency $\Gamma$ the Markovian results would be reproduced, one can recognize deviations from the Markovian separability times $\tau_1$ and $\tau_2$ due to non-Markovian and strong coupling effects. The non-weak system-environment interaction accelerates decoherence with the consequence that the initial entanglement caused by the squeezing is lost faster (fig.\ref{fig1}). In contrast to the Born-Markovian results in ref. \cite{prauzner} the initial entanglement is also lost if the critical value $\xi_c$ is exceeded. Specific non-Markovian effects (which are relevant if the bath correlation time scale $\Gamma^{-1}$ is comparable to the separability time scale) are shown in fig.\ref{fig2}: Non-Markovian influences prolong the separability time $\tau_s$ and may lead to a Gaussian decay of the logarithmic negativity instead of an exponential decay. Furthermore, fig.\ref{fig2} shows that partial revivals of quantum correlations can occur after an initial loss of entanglement. This can be seen from the oscillations of the logarithmic negativity between positive values (entanglement) and zero (separability). The same conclusions can be drawn from the criterion for marginal and global purities. The two mode state is separable if the total purity $\mu(t)$ is in the region determined by eq. \eqref{kritpur2} (gray shaded area in fig.\ref{fig1}-\ref{fig4}). \\
The stationary state in fig.\ref{fig1} and \ref{fig2} is separable in each case. However it is possible that quantum correlations exist in the stationary state. This becomes obvious from fig. \ref{fig3} and \ref{fig4}, which show examples for low temperatures. Despite a loss of entanglement at short times $t<\tau_s$ quantum correlations between the two modes emerge again at later times and persist even in the steady state.
Thus, in this model there is not just a separability or disentangling time $\tau_s$ but also a re-entangling time $\tau_e$ after which the two mode state remains entangled asymptotically. However it is difficult to find this characteristic time scale numerically due to the possible oscillations with partial revivals of the logarithmic negativity. Increasing the coupling strength $\gamma$ reduces the separability time $\tau_s$ as well as the re-entangling time $\tau_e$ as can be seen from a comparison between fig.\ref{fig3} and fig.\ref{fig4}.  The initial entanglement resulting from the squeezing is destroyed faster whereas environment-induced asymptotic quantum correlations are generated sooner. The re-entanglement time scale is strongly affected by the system bath interaction strength $\gamma$ but the initial squeezing with parameter $\xi$ has little effect on the stationary quantum correlations, in contrast to the results in previous works \cite{prauzner, an2}. Non-Markovian effects (induced by small cut-off frequency $\Gamma$) just extend the separability time scale $\tau_s$ but usually do not influence the re-entangling time $\tau_e$. Entanglement of the two modes can exist in the stationary state if the bath temperature is below a critical value, that depends on the coupling parameter $\gamma$ and the cut-off frequency $\Gamma$. \\
In the following we derive a simple separability criterion for the stationary state.
Since the normal coordinates $R=(q_1+q_2)/2$ and $p_x=(p_1-p_2)/2$ are EPR-like operators we can use product version \eqref{eprtan} of the EPR-separability criterion in the form
\begin{equation}\label{kritRx}
\langle R^2\rangle\langle p_x^2\rangle \geq
\frac{1}{16}\quad \Leftrightarrow \quad \langle
(q_1+q_2)^2\rangle\langle (p_1-p_2)^2\rangle \geq 1
\end{equation}
in order to derive a necessary condition for the existence of quantum correlations in the stationary state. Since in this case $\langle \{q_1,q_2\}\rangle <0$, $\langle \{ p_1,p_2\}\rangle>0$ and the stationary covariance matrix of the modified QBM model is given in symmetric standard form, the product criterion \eqref{eprtan} is equivalent to the PPT-criterion \cite{kimlee} and thus detects the same set of entangled covariance matrices. Using the stationary correlations in eq. \eqref{xxchi} and from \eqref{xrelax} this separability criterion can be rewritten in the form
\begin{equation}\label{kritRx2}
\frac{\omega_0}{\pi}\int_{0}^{\infty}d\omega\,(2\bar{n}+1)\coth(\frac{1}{2}\beta\hbar\omega)\,\mbox{Im}\{\tilde\chi(\omega)\}\geq\frac{1}{4},
\end{equation}
with mean occupation number $\bar{n}=(e^{\beta\hbar\omega_0}-1)^{-1}$ of the relative coordinate.
\begin{figure}[t]
\psfrag{x}[c]{$kT/\hbar\omega_0$} \psfrag{y}[c]{\small $E_{\mathcal N}$ vs. $I(\rho)$}
\psfrag{a}[c]{$I(\rho)$}\psfrag{b}[c]{$\langle \{p_1,p_2\}\rangle$}\psfrag{c}[c]{$\langle \{q_1,q_2\}\rangle$}\psfrag{d}[c]{$E_{Rx}$}\psfrag{e}[c]{$E_{\mathcal{N}}$}\psfrag{f}[c]{$S$}
\begin{center}
 \includegraphics[width=7.5cm,clip]{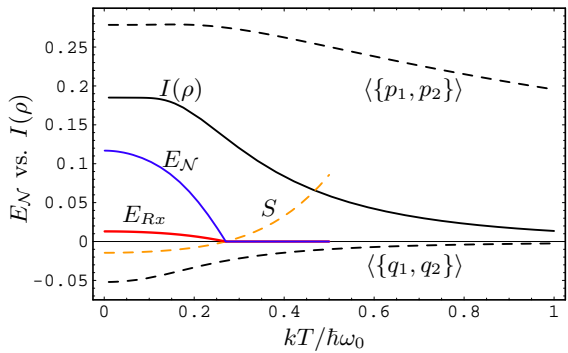}
 \caption[]{\label{fig5}\sl\small Separability criterion $S(\infty)$ of eq. \eqref{simkrit}, logarithmic negativity $E_{\mathcal{N}}$ \eqref{logNeg} and $E_{Rx}={\rm max}\{\frac{1}{16}-\langle R^2\rangle\langle p_x^2\rangle,0\}$ compared to the mutual information $I(\rho)$ \eqref{Imutualgauss} and the stationary correlations $\langle \{p_1,p_2\}\rangle$, $\langle \{q_1,q_2\}\rangle$ for the two mode state  in dependence of the temperature of the common reservoir. Parameters $\gamma=2\omega_0$ and $\Gamma=10\omega_0$.}
 \end{center}
\end{figure}\begin{figure}[t]
\begin{center}
\psfrag{t}[c]{\small $\Gamma/\omega_0$} \psfrag{x}[c]{\small$\gamma/\omega_0$} \psfrag{P}[c]{\small $E_{Rx}$}
 \includegraphics[width=7.5cm,clip]{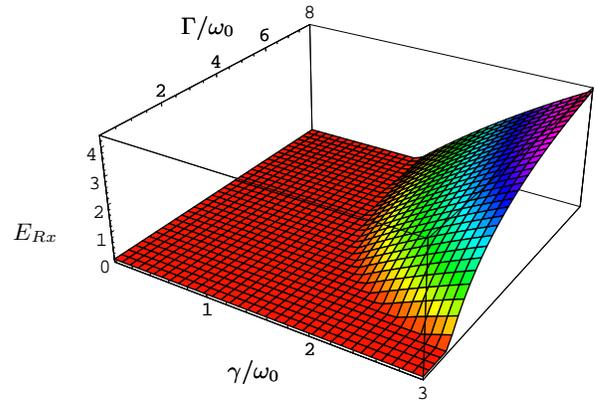}
 \caption[]{\label{fig6}\sl\small Stationary entanglement measured by the function $E_{Rx}={\rm max}\{\frac{1}{16}-\langle R^2\rangle\langle p_x^2\rangle,0\}$ ($\times$ $10^{-4}$) of a two-mode state in a common reservoir in dependence of coupling strength $\gamma$ and cut-off frequency $\Gamma$ at temperature $T=0.25\hbar\omega_0/k$.}
 \end{center}
\end{figure}
Figure \ref{fig5} compares this criterion in form of the function $E_{Rx}={\rm max}\{\frac{1}{16}-\langle R^2\rangle\langle p_x^2\rangle,0\}$  with the partial transposition separability criterion \eqref{simkrit} and the logarithmic negativity \eqref{logNeg} in dependence of the bath temperature. All these criteria give the same critical temperature. Also plotted are the stationary correlations $\langle \{q_1,q_2\}\rangle$ and $\langle \{p_1,p_2\}\rangle$ as well as the mutual information of a two mode Gaussian state which is defined by
\begin{equation}\label{Imutualgauss}
I(\rho)=S_v(\rho_1)+S_v(\rho_2)-S_v(\rho),
\end{equation}
where $S_v(\rho)$ is the von Neumann entropy of the total two mode state and $S_v(\rho_1)$, $S_v(\rho_2)$ are the von Neumann entropies of the reduced states $\rho_1=\mbox{Tr}_2[\rho]$ and $\rho_2=\mbox{Tr}_1[\rho]$. The von Neumann entropy of the reduced systems is equivalent to that of a single mode state and is given by
\begin{eqnarray}\label{SvN}
S_{v}(\rho_j)=\frac{1-\mu_j}{2\mu_j}\ln\frac{1+\mu_j}{1-\mu_j}-\ln\frac{2\mu_j}{1+\mu_j},
\end{eqnarray}
while the total entropy can be derived from the symplectic eigenvalues $\nu_{\pm}$ of $\mathbf{V}$ and reads \cite{werner2, serafini7}:
\begin{equation} S_v(\rho)=f(\nu_-)+f(\nu_+)\end{equation} with $
f(\nu_{\pm})=(\nu_{\pm}+\frac{1}{2})\ln(\nu_{\pm}+\frac{1}{2})-(\nu_{\pm}-\frac{1}{2})\ln(\nu_{\pm}-\frac{1}{2})$.
The mutual information is a measure of the amount of quantum and classical correlations \cite{zurekqd} and decays more slowly with temperature than the logarithmic negativity. Above a critical temperature (where the logarithmic negativity becomes zero) only classical correlations remain. The critical temperature depends on the spectral density of the bath and on the response function of the subsystem. The existence of stationary quantum correlations in dependence of the coupling constant $\gamma$ and the cut-off frequency $\Gamma$ is illustrated by figure \ref{fig6}. If the system-reservoir interaction constant becomes strong enough at a given temperature quantum correlations are still present in the stationary two-mode-state.

\section{summary and conclusions}
In this paper we have studied the dynamics of two mode squeezed states in a common thermal reservoir within an extended quantum Brownian motion model. Contrary to the two-reservoir model a common bath can provide an indirect coupling between the two subsystems and therefore a mechanism to correlate them. Thus, the coupling to the environment induces not only decoherence leading to separability but also can generate correlations resulting in asymptotic entanglement at low temperatures. The influence of non-Markovian and strong coupling effects on separability times scales was compared to previous results for Markovian dynamics. Non-Markovian effects increase the separability time but have little effect on the re-entangling time. The re-entangling time depends first of all on the system bath interaction strength. Increasing this coupling constant enforces entanglement generation and therefore reduces the re-entangling time. However, it also enhances decoherence and thus weakens the initial quantum correlations reducing the separability time.
For given parameters of the bath spectral density asymptotic entanglement is only present below a critical temperature. From the separability criterion for EPR-like operators a separability criterion was derived, which depends on the bath temperature and the response function of the open quantum system. Summarizing, the extended quantum Brownian motion model of a two-mode continuous variable system in a common reservoir provides an example of the case where environment-induced entanglement generation counteracts environment-induced decoherence. In view of the theory of quantum information with continuous variable systems these results may be relevant for experimental applications of correlated quantum channels. \\

{\sl Acknowledgement --} We would like to thank J. Eisert for valuable comments on the subject of this paper.

\end{document}